\documentclass[aps,twocolumn,showpacs,prl,groupedaddress,superscriptaddress,nofootinbib,longbibliography]{revtex4-2}
\usepackage{graphicx,amsmath,amssymb,amsbsy,subfigure,hyperref,bbm,times,txfonts}
\usepackage{color}
\usepackage{graphicx}
\usepackage{epstopdf}
\usepackage{amsmath}
\usepackage[marginal]{footmisc}
\usepackage{makecell}

\providecommand{\openone}{\leavevmode\hbox{\small1\kern-3.8pt\normalsize1}}
\hypersetup{
	colorlinks,%
	citecolor=blue,%
	linkcolor=blue,%
	urlcolor=blue
}

\providecommand{\fu}[1]{\textcolor{black}{#1}}

\makeatletter

\newcommand{\Rmnum}[1]{\expandafter\@slowromancap\romannumeral #1@}
\makeatother
\begin{document}
\title{\fu{Quantum Refrigeration via Valley-Selective {Bright-Dark} Excitonic Dynamics}}
\author{Zhichu Chen}
\thanks{These authors contributed equally to this work.}
\affiliation{School of Physics and Physical Engineering, Qufu Normal University, 273165, Qufu, China}
\author{Kang Lan}
\thanks{These authors contributed equally to this work.}
\affiliation{School of Physics and Physical Engineering, Qufu Normal University, 273165, Qufu, China}
\author{Yingtong Zhu}
\affiliation{School of Physics and Physical Engineering, Qufu Normal University, 273165, Qufu, China}
\author{Zhongxiao Man}
\affiliation{School of Physics and Physical Engineering, Qufu Normal University, 273165, Qufu, China}
 \author{Ning Hao}
\affiliation{Anhui Province Key Laboratory of Low-Energy Quantum Materials and Devices , High Magnetic Field Laboratory, HFIPS, Chinese Academy of Sciences, Hefei, Anhui 230031, China}
\author{Ping Zhang}
 \thanks{zhang\_ping@iapcm.ac.cn}
\affiliation{School of Physics and Physical Engineering, Qufu Normal University, 273165, Qufu, China}
\affiliation{Beijing Computational Science Research Center, Beijing 100084, China}
\author{Jiyong Fu}
\thanks{yongjf@qfnu.edu.cn}
\affiliation{School of Physics and Physical Engineering, Qufu Normal University, 273165, Qufu, China}
\affiliation{Instituto de F\'{\i}sica, Universidade de Bras\'{\i}lia, Bras\'{\i}lia-DF 70919-970, Brazil}

\begin{abstract}
  We establish valley excitons as a thermodynamic working medium for
continuous quantum refrigeration. \fu{We develop a microscopically grounded five-level open-system framework for monolayer MoS$_2$,
where opposite magnetic-field shifts of the bright--dark exciton
splittings in the $K$ and $K'$ valleys
render the corresponding transitions spectrally
distinguishable, enabling selective coupling to \emph{externally} controlled
hot and cold thermal photonic reservoirs.}
\fu{We further incorporate \emph{intrinsic} phonon-assisted bright--dark relaxation and show that its competition
with engineered photonic heat exchange leads to two distinct
thermodynamic regimes: strict refrigeration, where heat is extracted
from the cold photonic reservoir and released to the hot one, and a
phonon-dissipative regime, where both photonic reservoirs \emph{feed} the
excitonic system while energy is dissipated into the lattice.}
\fu{Coherent intervalley electron--hole exchange within
the dark-exciton sector redistributes the extracted energy between
valleys,} \fu{while magnetic field and optical polarization provide direct
control of the refrigeration response.}
  \fu{Our results uncover a distinct thermodynamic functionality of
valley excitons and establish a direct connection between valley
excitonics and quantum thermodynamics in two-dimensional
semiconductors.}
\end{abstract}

\maketitle

\textit{Introduction}.---
Monolayer transition-metal dichalcogenides (TMDCs) $MX_2$
($M$=Mo, W; $X$=S, Se) host two inequivalent valleys at the
$K$ and $K'$ points of the Brillouin zone, selectively
addressable by circularly polarized (CP) excitation
\cite{TMDC1,TMDC2,TMDC8,TMDC6,TMDC4,TMDC3,TMDC5,TMDC7}.
Their \textit{spin-valley-locked} electronic structure
\cite{WOS:000546940100012,Zhang2023} and strong many-body
interactions provide a platform for controlling valley-exciton
coherence
\cite{PhysRevLett.123.096803,Qiu2019,PhysRevLett.121.116102,
PhysRevB.108.035419} and generating entangled states
\cite{PhysRevB.92.075409,PhysRevB.107.035404}.
\fu{Beyond quantum-state control, valley-selective excitation
and internal excitonic dynamics offer a route to manipulating
energy flow, raising the question of whether valley excitons
can serve as a thermodynamic working medium for heat extraction
and refrigeration.}

\fu{This question is particularly timely because many existing
quantum refrigeration schemes employ idealized few-level
or harmonic-oscillator models}
\cite{PhysRevResearch.2.023120,PhysRevE.91.050102,PhysRevA.104.042208},
\fu{with coherent dynamics}
\cite{PhysRevX.5.031044,Hammam_2021,QTM7},
\fu{engineered level spacings}
\cite{QTM4,PhysRevE.98.012117},
\fu{and controlled coupling to multiple reservoirs}
\cite{QTM2,QTM3}.
\fu{Implementing these ingredients in solid-state materials
while accounting for intrinsic relaxation remains challenging,
and experimental demonstrations have largely remained
at the proof-of-principle stage}
\cite{Baugh2005,Maslennikov2019,Aamir2025,PhysRevLett.132.210403}.
Valley excitons in monolayer TMDCs offer internal dynamical
channels \cite{TMDC9,PhysRevB.100.115306},
strong light--matter coupling \cite{Luo2024},
and electromagnetic tunability
\cite{PhysRevB.110.125420,PhysRevB.108.035419},
\fu{providing a material setting in which intrinsic excitonic
dynamics can be harnessed for engineered heat exchange.
However, a microscopic framework establishing whether these
ingredients support quantum refrigeration in the presence
of intrinsic lattice relaxation remains lacking.}

\begin{figure}[h]
\vskip -0.45cm
\begin{center}
\setlength{\abovecaptionskip}{0.cm}
{\includegraphics[width=\linewidth]{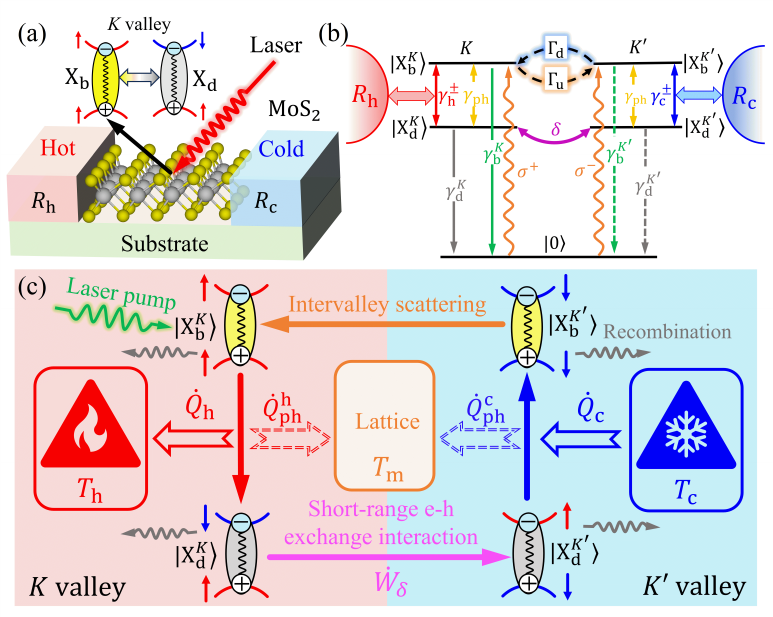}}
\end{center}
\vskip -0.6cm
\caption{(Color online) \fu{(a) Valley-excitonic working medium in monolayer MoS$_2$
coupled to hot ($R_{\mathrm h}$) and cold ($R_{\mathrm c}$)
thermal photonic reservoirs.
(b) Five-level scheme with photonic absorption/emission
rates $\gamma_{\mathrm {h,c}}^{\pm}$, intrinsic phonon-assisted
bright--dark rates $\gamma_{\mathrm{ph}}$, bright-exciton
intervalley rates $\Gamma_{\mathrm {u,d}}$, recombination rates
$\gamma_{\mathrm {b,d}}^{\xi}$, and coherent dark-exciton
exchange strength $\delta$.
(c) Energy-transfer pathways:
$\dot Q_{\mathrm {h,c}}$ are photonic heat currents,
$\dot Q_{\mathrm{ph}}^{\mathrm {h,c}}$ are phonon heat currents
between the two valleys and the common lattice at $T_m$,
and $\dot W_\delta$ is the exchange-associated work current.}}
\vskip -0.45cm
\label{fig1}
\end{figure}

Here we establish valley excitons as a thermodynamic working
medium for continuous quantum refrigeration
\fu{in monolayer MoS$_2$.
Opposite magnetic-field shifts of the bright--dark exciton
splittings in the $K$ and $K'$ valleys render the corresponding
transitions spectrally distinguishable, enabling selective
coupling to externally controlled hot and cold thermal
photonic reservoirs.}
We develop a \fu{microscopically grounded} five-level
open-system framework and \fu{derive a compact criterion
for cold-reservoir heat extraction governed by the
bright-to-dark population ratio in the cold-coupled valley.
Competition between engineered photonic heat exchange and
intrinsic phonon-assisted bright--dark relaxation produces
two distinct operating regimes: strict refrigeration,
in which heat is extracted from the cold photonic reservoir
and released into the hot one, and a phonon-dissipative regime,
in which both photonic reservoirs supply energy while
phonon relaxation dissipates energy into the lattice.
Coherent intervalley electron--hole (e--h) exchange within
the dark-exciton sector redistributes energy between valleys,
while magnetic field and optical polarization control
the refrigeration response.
These results connect valley excitonics with quantum
thermodynamics and identify the microscopic processes
governing refrigeration in two-dimensional semiconductors.}

\textit{Valley-selective refrigeration mechanism and dynamics}.---
The relevant five-level subsystem in monolayer MoS$_2$
comprises bright ($|\mathrm X_{\mathrm b}^{\xi}\rangle$)
and dark ($|\mathrm X_{\mathrm d}^{\xi}\rangle$) excitons
in the $\xi=K,K'$ valleys and the ground state $|0\rangle$
reached after recombination
\cite{PhysRevLett.123.096803,PhysRevLett.129.067402}.
\fu{The excitons couple to externally controlled hot and cold
thermal photonic reservoirs, $R_{\mathrm h}(T_\mathrm{h})$ and
$R_{\mathrm c}(T_\mathrm{c})$, and a common lattice environment
at $T_\mathrm{m}$ [Fig.~\ref{fig1}(a)].
Opposite magnetic-field shifts of the bright--dark splittings
spectrally separate the two valley transitions, allowing
selective coupling of $K$ and $K'$ to $R_{\mathrm h}$ and
$R_{\mathrm c}$, respectively.
Photonic heat exchange competes with intrinsic phonon-assisted
bright--dark relaxation, while bright-exciton intervalley
scattering and coherent dark-exciton exchange connect the
valleys [Figs.~\ref{fig1}(b) and \ref{fig1}(c)].}
We consider the total Hamiltonian
$H_{\mathrm{tot}}=H_S+H_R+H_I$
\cite{PhysRevB.108.035419,PhysRevB.110.125420},
where $H_S$, $H_R$, and $H_I$ describe the excitons,
environments, and their couplings:
\begin{align}
H_S={}&H_X+H_{XL}+\fu{H_\delta+H_J},\nonumber\\
H_R={}&\textstyle\sum_{\mu=\mathrm h,\mathrm c}\sum_l
\hbar\omega_{\mu,l}a_{\mu,l}^{\dagger}a_{\mu,l}
+\fu{\textstyle\sum_{\mathbf q}\hbar\omega_{\mathbf q}
b_{\mathbf q}^{\dagger}b_{\mathbf q}},\nonumber\\
H_I={}&\fu{\textstyle\sum_l
\left(\kappa_{\mathrm h,l}\sigma_{\mathrm{bd}}^K a_{\mathrm h,l}
+\kappa_{\mathrm c,l}\sigma_{\mathrm{bd}}^{K'}a_{\mathrm c,l}
\right)}\nonumber\\
&\fu{+\textstyle\sum_{\xi,\mathbf k,\mathbf q}
\lambda_{\xi,\mathbf k,\mathbf q}
c_{\xi\mathrm d,\mathbf k}^{\dagger}
c_{\xi\mathrm b,\mathbf k+\mathbf q}
b_{\mathbf q}^{\dagger}}+\rm{H.c.},
\label{eqs1}
\end{align}
with $\sigma_{ij}=|i\rangle\langle j|$.
Here $a_{\mu,l}$ annihilates a photon in reservoir $R_\mu$,
\fu{$b_{\mathbf q}$ a lattice phonon, and
$c_{\xi s,\mathbf k}$ an exciton in valley $\xi$,
branch $s=\mathrm b,\mathrm d$, and 
momentum $\mathbf k$.}
\fu{The coupling strengths $\kappa_{\mu,l}$ and $\lambda_{\xi,\mathbf{k},\mathbf{q}}$ describe the engineered photon-assisted and intrinsic phonon-assisted bright--dark transitions, respectively.}
\fu{Spin-orbit coupling (SOC) induces spin admixture
that enables bright--dark conversion
\cite{WangLi2008,Jiang2021,Lan2026}.
The engineered photonic and intrinsic phononic channels
arise from distinct electromagnetic and exciton--phonon
interactions (SM, Secs.~IV and V).
Photon-assisted spin flips in quantum dots
\cite{Braakman2014} and photonic control of TMDC dark
excitons \cite{Lo2022,Shan2022} provide related routes
to accessing spin-forbidden transitions.}
 The excitonic terms are
\begin{align}
H_X={}&\textstyle\sum_\xi
\left(E_{\mathrm b}^{\xi}\sigma_{\mathrm{bb}}^{\xi}
+E_{\mathrm d}^{\xi}\sigma_{\mathrm{dd}}^{\xi}\right),
\nonumber\\
H_\delta={}&\fu{({\delta}/{2})
\left(\sigma_{\mathrm{dd}}^{KK'}
+\sigma_{\mathrm{dd}}^{K'K}\right)},\nonumber\\
H_J={}&\fu{\textstyle\sum_{\mathbf k}J(\mathbf k)
\left[\sum_\xi c_{\xi\mathrm b,\mathbf k}^{\dagger}
c_{\xi\mathrm b,\mathbf k}
-\left(e^{-2i\phi_{\mathbf k}}
c_{K\mathrm b,\mathbf k}^{\dagger}
c_{K'\mathrm b,\mathbf k}+\mathrm{H.c.}\right)\right]},
\nonumber\\
H_{XL}={}&\hbar\left(
\Omega_K e^{i\omega_Rt}\sigma_{0\mathrm b}^{K}
+\Omega_{K'}e^{i\omega_Lt}\sigma_{0\mathrm b}^{K'}
\right)+\mathrm{H.c.},
\label{eqs2}
\end{align}
where
\fu{$\sigma_{\mathrm{dd}}^{\xi\xi'}=|\mathrm X_{\mathrm d}^{\xi}\rangle \langle\mathrm X_{\mathrm d}^{\xi'}|$, and the optical amplitudes $\Omega_K$ ($\Omega_{K'}$)
and frequencies $\omega_R$ ($\omega_L$) correspond
to right $\sigma^+$ (left $\sigma^-$) CP excitation.
Short-range (SR) e--h exchange $H_\delta$ coherently couples the dark-valley states.
Long-range (LR) exchange $H_J$ produces an effective valley exchange field dependent
on the exciton momentum $\mathbf k$,
with $J(\mathbf k)$ its amplitude and $\phi_{\mathbf k}$
the polar angle of $\mathbf k$. Momentum scattering randomizes this field; averaging
over the momentum dynamics yields the effective bright-exciton intervalley dissipator (SM, Sec.~II).}
Transforming the remaining Hamiltonian with
$U=\exp(i\bar Ht/\hbar)$, 
$\bar H=\hbar(\omega_R\sigma_{\mathrm{bb}}^K+
\omega_L\sigma_{\mathrm{bb}}^{K'})$
\cite{PhysRevLett.98.240601,PhysRevResearch.2.043187,
PhysRevE.101.062121},
gives $\mathcal H_S=H_X-\bar H+H_\delta+H_{XL}^{R}$, with $H_{XL}^{R}=\hbar(\Omega_K\sigma_{0\mathrm b}^{K}+\Omega_{K'}\sigma_{0\mathrm b}^{K'})+\mathrm{H.c.}$
Within the weak-coupling approximation, \fu{the reduced density matrix obeys}
\begin{equation}
\fu{\dot\rho=-({i}/{\hbar})[\mathcal H_S,\rho]
+\left(\mathcal L_{\mathrm f}+\mathcal L_{\mathrm r}
+\mathcal L_{\mathrm{ph}}
+\textstyle{\sum_{\mu=\mathrm h,\mathrm c}}\mathcal L_\mu\right)\rho.}
\label{eqs3}
\end{equation}
\fu{Here $\mathcal L_{\mathrm f}$ describes bright-exciton
intervalley transfer through exchange-induced mixing
and phonon-assisted transitions
\cite{Surrente2018,liang2024selective,Castro2023Intervalley}.
The phonon-assisted rates satisfy Boltzmann detailed
balance at $T_\mathrm{m}$ (SM, Secs.~III and VI).
The remaining dissipators describe recombination
($\mathcal L_{\mathrm r}$), intrinsic phonon-assisted
bright--dark relaxation ($\mathcal L_{\mathrm{ph}}$),
and photonic bright--dark transitions ($\mathcal L_\mu$).
Their explicit forms are given in the SM, Sec.~VI.}
\fu{Equation~\eqref{eqs3} describes momentum-integrated
exciton populations. Rapid intrabranch momentum relaxation
is considered to establish a thermal distribution at $T_\mathrm{m}$
\cite{Chan2025,Werner2026}, while the valley-resolved
bright- and dark-exciton populations remain nonequilibrium
dynamical variables~\cite{Zhu2014Kerr,Glazov2015Valley,Kumar2025Valley}.
Averaging the momentum-dependent radiative rate over
this distribution, with direct emission restricted
to the light cone, yields the effective
exciton recombination rate
\cite{life,TMDC3} (SM, Sec.~VI).}

\textit{Quantum refrigeration criterion}.---
\fu{Using $H_X$ to evaluate the excitonic energy, we define
the photonic heat current as positive from reservoir
$R_\mu$ into the  excitonic working medium:}
\begin{equation}
\dot Q_\mu
=\mathrm{Tr}(H_X[\mathcal L_\mu\rho])
=\Delta E_{\mathrm{bd}}^\xi\Gamma_0\bar n_\mu
\left(\rho_{\mathrm d}^\xi
-e^{\Delta E_{\mathrm{bd}}^\xi/k_BT_\mu}
\rho_{\mathrm b}^\xi\right),
\label{eqs4}
\end{equation}
\fu{where $\mu=\mathrm h,\mathrm c$ corresponds to
$\xi=K,K'$, respectively.}
Here $\Delta E_{\mathrm{bd}}^\xi
=E_{\mathrm b}^\xi-E_{\mathrm d}^\xi>0$ is the bright--dark
splitting, and $\rho_{\mathrm {b,d}}^\xi$ are the exciton
populations.
\fu{The photon occupation
$\bar n_\mu=[\exp(\Delta E_{\mathrm{bd}}^\xi/k_BT_\mu)-1]^{-1}$
determines the absorption and emission rates,
$\gamma_\mu^+=\Gamma_0\bar n_\mu$ and
$\gamma_\mu^-=\Gamma_0(\bar n_\mu+1)$.
The reservoir spectral rates are
$\Gamma_\mu(\omega)=(2\pi/\hbar^2)\sum_l
|\kappa_{\mu,l}|^2\delta(\omega-\omega_{\mu,l})$
\cite{PhysRevE.85.061126}.
For simplicity, we take $\Gamma_{\mathrm h}(\omega_{\mathrm{bd}}^K)\simeq\Gamma_{\mathrm c}(\omega_{\mathrm{bd}}^{K'})\equiv\Gamma_0$, where
$\omega_{\mathrm{bd}}^\xi=\Delta E_{\mathrm{bd}}^\xi/\hbar$ (SM, Sec.~V).
\fu{The zero-field bright--dark splitting of approximately
$14$ meV \cite{MoS2} corresponds to $3.4$ THz.}
The intrinsic phonon current $\dot Q_{\mathrm{ph}}=\mathrm{Tr}(H_X\mathcal L_{\mathrm{ph}}\rho)$
is evaluated separately (SM, Sec.~VIII).
Exciton energies and material parameters are given in the
SM, Secs.~I and VII.}

To extract heat from the cold photonic reservoir,
a positive cooling power, $\dot Q_{\mathrm c}>0$,
is required. From Eq.~\eqref{eqs4}, we obtain a compact
\fu{cold-reservoir heat-extraction criterion},
\begin{equation}
\exp(-\Delta E_{\mathrm{bd}}^{K'}/k_BT_{\mathrm c})
>
\rho_{\mathrm b}^{K'}/\rho_{\mathrm d}^{K'}.
\label{eq:condition}
\end{equation}
Equation~\eqref{eq:condition} makes the mechanism transparent:
heat extraction requires the cold-coupled valley to accumulate
a sufficiently large dark-exciton population relative to
the bright-exciton population.
\fu{This population imbalance favors photon absorption
over emission and therefore provides a direct criterion
for controlling cold-reservoir heat extraction.
Strict refrigeration further requires heat release into
the hot photonic reservoir, $\dot Q_{\mathrm h}<0$.}

\begin{figure}
\begin{center}
\setlength{\abovecaptionskip}{0.cm}
{\includegraphics[width=\linewidth]{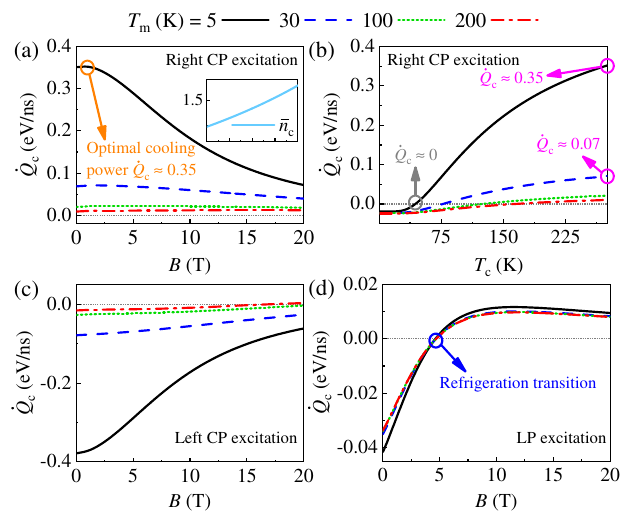} }
\end{center}
\vskip -0.6cm
\caption{(Color online) Control and polarization selectivity
\fu{of cooling in the photonic-channel-dominated limit}.
(a), (b) Cooling power $\dot Q_{\mathrm c}$ versus
\fu{magnetic field $B$ (a) and cold-reservoir temperature
$T_{\mathrm c}$ (b)}  under right CP excitation.
In (a), \fu{the orange circle marks the optimal cooling
power; the inset shows the photon occupation
$\bar n_{\mathrm c}$.} In (b), pink (gray) circle indicates $\dot{Q}_{\mathrm c}$ at $T_{\mathrm c}=275$ K (44 K).
(c), (d) Cooling power $\dot Q_{\mathrm c}$ versus
magnetic field under left CP excitation (c) and
LP excitation (d).
In all panels, $T_{\mathrm m}=5$, $30$, $100$, and $200$ K.}
\vskip -0.6cm
\label{fig2}
\end{figure}

\textit{Control and polarization selectivity \fu{in the photonic-channel-dominated limit}}.---
\fu{We first consider the limit in which intrinsic
phonon-assisted bright--dark relaxation is weak compared
with the engineered photonic transitions. This limit isolates
the valley-selective photonic heat-exchange mechanism and
provides a reference for the phonon--photon competition
discussed below.}
To promote heat extraction from $R_{\mathrm c}$ according
to Eq.~\eqref{eq:condition}, we exploit the valley-selective
optical selection rules of monolayer MoS$_2$
\cite{WOS:000546940100012}.
Right CP excitation ($\sigma^+$) predominantly generates
bright excitons in the $K$ valley. Intervalley scattering
transfers these excitons into $K'$, where relaxation into
the lower-energy dark state builds the population imbalance
required for cooling.

By tuning the magnetic field, we obtain an optimal cooling
power of $0.35$ eV/ns at $T_\mathrm{m}=5$ K in the low-field region
($B<2$ T), marked by the orange circle in
Fig.~\ref{fig2}(a).
\fu{The low-field response reflects two competing effects:
increasing $B$ raises the photon occupation $\bar n_{\mathrm c}$
[inset of Fig.~\ref{fig2}(a)], favoring photon absorption
from the cold reservoir, while reducing the bright--dark
splitting $\Delta E_{\mathrm{bd}}^{K'}$ and hence the energy
extracted per absorbed photon (SM, Sec.~I and Fig.~S1).
Their competition produces a nearly constant cooling
power at low fields.}
At higher fields ($B\gtrsim2$ T), the valley splitting
suppresses $K\rightarrow K'$ intervalley transfer,
reducing the dark-exciton population in the cold-coupled
valley (\fu{SM, Sec.~X}).
Together with the decreasing $\Delta E_{\mathrm{bd}}^{K'}$,
this leads to a rapid decay of $\dot Q_{\mathrm c}$.

The cold-reservoir temperature $T_{\mathrm c}$ provides
a second thermal control knob [Fig.~\ref{fig2}(b)].
At low $T_{\mathrm c}$, Eq.~\eqref{eq:condition} is not
satisfied and no cooling occurs.
\fu{Increasing $T_{\mathrm c}$ raises the photon occupation
$\bar n_{\mathrm c}$, promoting dark-to-bright absorption
in the $K'$ valley and driving the onset of cooling
at $\dot Q_{\mathrm c}=0$.}
The enhancement is strongest at low material temperature:
the cooling power at $T_\mathrm{m}=5$ K exceeds that at $30$ K
by nearly a factor of five.
By contrast, increasing $T_\mathrm{m}$ suppresses cooling
[Figs.~\ref{fig2}(a) and \ref{fig2}(b)].
\fu{The longer effective bright-exciton radiative lifetime
increases the bright population, while thermal phonons
weaken the relative asymmetry between upward and downward
intervalley scattering rates (SM, Secs.~III and VI).
Both effects hinder the population imbalance required
for cooling.} \fu{To establish the complete refrigeration cycle, we also
evaluate the heat current into the hot reservoir.
Throughout the cooling region under right CP excitation,
$\dot Q_{\mathrm h}<0$, showing that cold-reservoir heat
extraction is accompanied by heat release into
$R_{\mathrm h}$ (SM, Sec.~IX and Fig.~S2).}

A distinctive feature of the refrigeration response is its
polarization selectivity.
Under left CP excitation ($\sigma^-$), only a nearly vanishing
cooling signal is obtained over the magnetic-field range
considered [Fig.~\ref{fig2}(c)].
\fu{Direct excitation of the cold-coupled $K'$ bright branch
increases $\rho_{\mathrm b}^{K'}$, hindering the population
imbalance required by Eq.~\eqref{eq:condition}.}
Linearly polarized (LP) excitation pumps both valleys and
yields intermediate behavior [Fig.~\ref{fig2}(d)]:
\fu{positive $\dot Q_{\mathrm c}$ emerges only above a
threshold field and remains substantially weaker than
under right CP excitation.
As $B$ increases, the lower exciton energies in $K$ favor
intervalley redistribution toward that valley, reducing
the $K'$ bright population and allowing the heat-extraction
criterion to be satisfied.
The polarization contrast thus directly links the
refrigeration response to optical control of the
bright-to-dark population ratio in the cold-coupled valley.}

\begin{figure}
\begin{center}
\setlength{\abovecaptionskip}{0.cm}
{\includegraphics[width=\linewidth]{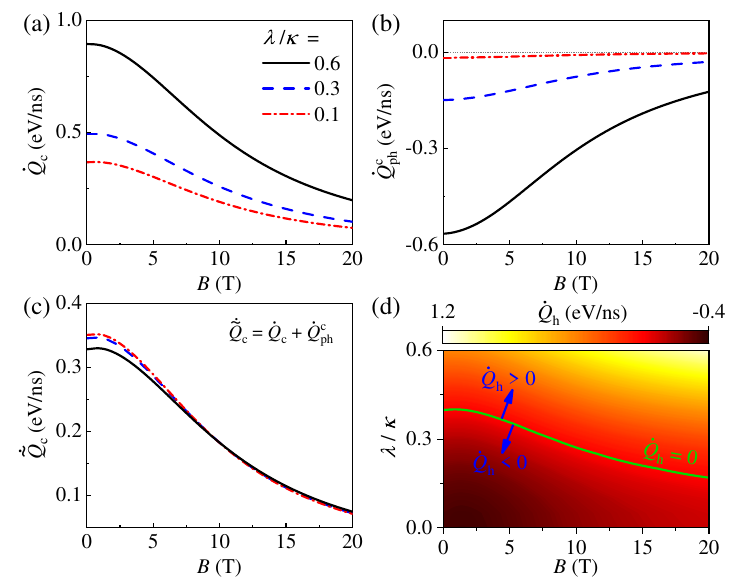} }
\end{center}
\vskip -0.6cm
\caption{(Color online) \fu{(a)--(c) Magnetic-field dependence of the cold-photonic
current $\dot Q_{\mathrm c}$ (a), the $K'$-valley phonon
current $\dot Q_{\mathrm{ph}}^{\mathrm c}$ (b), and their sum
$\dot{\tilde{Q}}_{\mathrm c}=\dot Q_{\mathrm c}+\dot Q_{\mathrm{ph}}^{\mathrm c}$ (c)
for $\lambda/\kappa=0.6$, $0.3$, and $0.1$, with
$\kappa=0.4$ THz.
(d) Hot-photonic current $\dot Q_{\mathrm h}$ versus
$B$ and $\lambda/\kappa$.
The green contour $\dot Q_{\mathrm h}=0$ separates
strict refrigeration
($\dot Q_{\mathrm c}>0,\dot Q_{\mathrm h}<0$)
from phonon-dissipative operation
($\dot Q_{\mathrm c}>0,\dot Q_{\mathrm h}>0$).}
}
\vskip -0.6cm
\label{fig3}
\end{figure}

\textit{\fu{Intrinsic phonon relaxation and thermodynamic crossover}}.---
\fu{We now examine the competition between intrinsic
phonon-assisted bright--dark relaxation and engineered
photonic heat exchange.
For simplicity, we approximate
$\lambda_{\xi,\mathbf k,\mathbf q}$ in Eq.~\eqref{eqs1}
by an  effective averaged constant $\lambda$ and characterize the relative
coupling strengths by $\lambda/\kappa$
(SM, Secs.~IV and VI).
For a given material, the photonic channel can be
controlled through the electromagnetic environment
and reservoir occupations~\cite{Liu2024,Vaaranta2022}.}
\fu{Figures~\ref{fig3}(a) and \ref{fig3}(b) show the
cold-reservoir current $\dot Q_{\mathrm c}$ and the
phonon current $\dot Q_{\mathrm{ph}}^{\mathrm c}$,
defined as positive from the lattice into the $K'$ valley
(SM, Sec.~VIII).
Increasing $\lambda/\kappa$ enhances
$\dot Q_{\mathrm c}$ while making
$\dot Q_{\mathrm{ph}}^{\mathrm c}$ more negative.
At low $T_\mathrm{m}$, phonon emission populates the lower-energy
dark state, favoring photon absorption from $R_{\mathrm c}$
while releasing energy into the lattice.
To quantify the balance between these pathways,
we evaluate their combined energy current,
\begin{equation}
\dot{\widetilde{Q}_{\mathrm c}}
=\dot Q_{\mathrm c}+\dot Q_{\mathrm{ph}}^{\mathrm c}.
\label{eq:qtilde}
\end{equation}
Despite the increase in $\dot Q_{\mathrm c}$,
$\dot{\widetilde{Q}_{\mathrm c}}$ remains nearly unchanged
[Fig.~\ref{fig3}(c)]:
the additional heat absorbed from $R_{\mathrm c}$
is largely released into the lattice through phonon
emission, leaving the net energy input to the $K'$-valley
excitons through these two channels nearly unchanged.}

\fu{The hot-side current reveals two distinct thermodynamic
regimes [Fig.~\ref{fig3}(d)].
At weak phonon coupling, $\dot Q_{\mathrm c}>0$ and
$\dot Q_{\mathrm h}<0$ define strict refrigeration:
heat is extracted from the cold photonic reservoir
and released into the hot one.
Stronger phonon coupling can reverse $\dot Q_{\mathrm h}$
while maintaining $\dot Q_{\mathrm c}>0$, producing
a phonon-dissipative regime in which both photonic
reservoirs supply energy while phonon relaxation
dissipates energy into the lattice.
Within the region $\dot Q_{\mathrm c}>0$, the contour
$\dot Q_{\mathrm h}=0$ separates these two regimes.
Intrinsic relaxation can thus enhance cold-reservoir
heat extraction while changing the thermodynamic
operating regime.
Results for left CP and LP excitation are given
in the SM (Sec.~XII).}

\begin{figure}
\begin{center}
\setlength{\abovecaptionskip}{0.cm}
{\includegraphics[width=\linewidth]{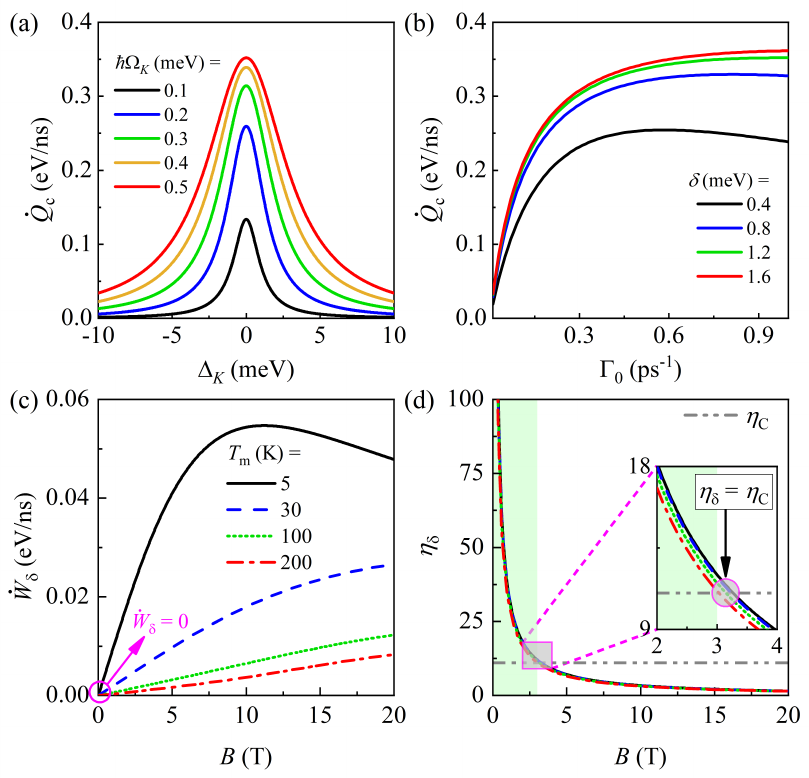} }
\end{center}
\vskip -0.6cm
\caption{(Color online) (a), (b) Cooling power $\dot Q_{\mathrm c}$ versus
laser detuning $\Delta_K$ (a) and \fu{photonic transition rate} $\Gamma_0$ (b).
In (a), $\hbar\Omega_K=0.1$--$0.5$ meV; in (b), $\delta=0.4$--$1.6$ meV.
(c), (d) \fu{Exchange-associated work current}
$\dot W_\delta$ (c) and \fu{exchange-based effective COP $\eta_\delta$} (d)
versus magnetic field for $T_m=5$, $30$, $100$, and $200$ K.
In (d), the gray double-dotted line marks the conventional
Carnot COP $\eta_{\mathrm C}$ for a standard two-terminal
power-driven refrigerator~\cite{PhysRevResearch.2.043302},
\fu{shown as a reference rather than a bound on
$\eta_\delta$}; inset: magnified view near
$\eta_\delta=\eta_{\mathrm C}$.
}
\vskip -0.6cm
\label{fig4}
\end{figure}

\textit{Laser detuning, internal channels, and work current}.---
\fu{Having established the phonon-induced thermodynamic
crossover, we return to the photonic-channel-dominated
limit to isolate how optical driving and coherent
dark-exciton exchange control refrigeration.}
In addition to magnetic field, reservoir temperature,
and excitation polarization, the refrigeration response
can be tuned by the laser detuning $\Delta_K$, defined
as the energy offset between the $K$-valley bright exciton
and the incident photon.
Figure~\ref{fig4}(a) shows that the cooling power is
maximized at resonant excitation, $\Delta_K=0$.
\fu{Resonant pumping maximizes the exciton population
and promotes population transfer into the cold-coupled
$K'$ dark branch (SM, Sec.~XI and Fig.~S4).}
Increasing the optical coupling strength $\hbar\Omega_K$
further enhances $\dot Q_{\mathrm c}$ at resonance.
Laser detuning and excitation strength therefore provide
additional optical controls of the refrigeration response.

The refrigeration mechanism is further governed by
\fu{photonic bright--dark transitions, characterized
by the vacuum rate $\Gamma_0$, and coherent coupling
between dark-valley states, governed by $\delta$}
[Fig.~\ref{fig4}(b)].
Although these channels play different roles, both
regulate the steady-state populations and hence the
cooling power.
Increasing $\delta$ enhances refrigeration,
\fu{with the enhancement gradually saturating
for $\delta\gtrsim1.6$ meV.}
Increasing $\Gamma_0$ further boosts $\dot Q_{\mathrm c}$,
particularly at larger $\delta$.
\fu{Here $\Gamma_0$ controls photon absorption and
emission between bright and dark excitons, while
$\delta$ connects the dark states in opposite valleys.
Their combined action links reservoir heat exchange
to internal exciton transfer and controls the population
imbalance required by Eq.~\eqref{eq:condition}.}

The exchange interaction constitutes a coherent internal
channel, distinct from heat exchange with the reservoirs.
\fu{Taking $H_X$ as the reference Hamiltonian for the
excitonic energy balance, we define its coherent
contribution as the exchange-associated work current,}
\begin{equation}
  \label{eq:wdelta}
\dot W_\delta=(i/\hbar)\mathrm{Tr}\!\left([H_\delta,H_X]\rho\right)=({\delta}/{\hbar})
\left(E_{\mathrm d}^{K'}-E_{\mathrm d}^{K}\right)
\operatorname{Im}\rho_{\mathrm{d}}^{KK'},
\end{equation}
where
$\rho_{\mathrm{d}}^{KK'}=\langle X_{\mathrm d}^{K}|\rho|X_{\mathrm d}^{K'}\rangle$
is the dark-valley coherence (SM, Sec.~VIII).
Unlike work channels introduced through external
coherent driving or generalized coherence resources
\cite{Schrauwen2025,Rodrigues2024,Watanabe2025},
this contribution arises intrinsically from coherent
exchange within the valley-excitonic subsystem.
In this sense, $\dot W_\delta$ measures
\fu{the exchange-mediated redistribution of bare-exciton
energy within the working medium, rather than an
independently supplied external power.}
Figure~\ref{fig4}(c) shows that $\dot W_\delta$
exhibits a pronounced magnetic-field dependence.
\fu{Equation~\eqref{eq:wdelta} identifies its two
essential ingredients: dark-valley energy splitting
and intervalley coherence.
At zero field, the dark states are degenerate,
so $[H_\delta,H_X]=0$ and $\dot W_\delta=0$,
irrespective of their population imbalance.
A finite field lifts this degeneracy and permits
a nonzero exchange-associated work current.}
At low material temperature ($T_\mathrm{m}=5$ K) and 
higher field ($B\gtrsim10$ T), suppressed bright-exciton
intervalley scattering reduces the population feeding
the $K'$ dark state.
This depletion weakens the exchange channel and leads
to saturation and eventual reduction of $\dot W_\delta$
[black curve in Fig.~\ref{fig4}(c)].

To assess the role of this internal channel, we define
an \emph{effective} coefficient of performance (COP),
$\eta_\delta={\dot Q_{\mathrm c}}/{\dot W_\delta}$,
which compares heat extracted from the cold reservoir
with the exchange-associated work current.
It excludes the external optical drive and therefore
\fu{characterizes the exchange channel rather than
the full thermodynamic COP of the refrigerator.}
As shown in Fig.~\ref{fig4}(d), $\eta_\delta$ can exceed
the conventional two-reservoir Carnot reference,
$\eta_{\mathrm C}=T_{\mathrm c}/(T_{\mathrm h}-T_{\mathrm c})$,
in the low-field regime.
\fu{This does not imply a violation of the Carnot bound:
the denominator here is an internal exchange contribution,
not the total external input power.}
The significance of $\eta_\delta$ is therefore
\emph{diagnostic}: it quantifies cold-reservoir heat
extraction relative to the internal exchange contribution.
As the field increases, $\eta_\delta$ decreases toward
the conventional reference
\fu{as the cooling power falls relative to
  $\dot W_\delta$.}

\textit{\fu{Experimental implementation scheme.---}}
\fu{We propose coupling an hBN-encapsulated MoS$_2$
monolayer to $R_{\mathrm h}(T_{\mathrm h})$ and
$R_{\mathrm c}(T_{\mathrm c})$ using two THz resonators
to concentrate the reservoir--exciton couplings
near the target transition frequencies.
Their modes are tuned respectively to the magnetically
separated $K$- and $K'$-valley bright--dark transitions
near $3.4$ THz \cite{MoS2}.
Continuous CP excitation drives the monolayer,
while its lattice remains anchored at $T_{\mathrm m}$.
Experimental MoS$_2$--THz resonator integration
\cite{Ji2019} provides a device precedent,
while cryogenic resonance tuning \cite{Jarc2022}
offers frequency matching to the valley transitions.
Resonator--reservoir connections have been demonstrated
in microwave quantum heat valves \cite{Ronzani2018},
and frequency-dependent reservoir coupling provides
control of mode occupations \cite{Bacciconi2026}.
The implementation requires resolved valley transitions
and thermal photon occupations maintained by the
respective reservoirs.
Refrigeration would manifest as net THz absorption
from $R_{\mathrm c}$ and emission into $R_{\mathrm h}$,
with excitation helicity controlling the valley-selective
response.}

\textit{Concluding remarks}.---
\fu{We have established valley excitons as a thermodynamic
working medium for continuous quantum refrigeration,
connecting valley excitonics with quantum thermodynamics.
In monolayer MoS$_2$, opposite magnetic-field shifts
of the bright--dark splittings enable valley-selective
coupling to externally controlled hot and cold thermal
photonic reservoirs.
A compact population criterion governs cold-reservoir
heat extraction, while coherent dark-exciton exchange
redistributes energy between valleys and magnetic field
and optical polarization control the refrigeration response.
Crucially, competition between engineered photonic
heat exchange and intrinsic phonon-assisted relaxation
produces a crossover between strict refrigeration
and phonon-dissipative operation, identifying the
conditions for cold-to-hot heat transfer and for
energy flow from both photonic reservoirs into the lattice.
Our results establish a thermodynamic functionality
of valley excitons and provide a framework for harnessing
optical selection rules, coherent exchange, and engineered
photonic environments to control heat flow in
two-dimensional semiconductors.}

\textit{Acknowledgments.} We wish to acknowledge the late Fanyao Qu, whose original insights, shared during discussions on incorporating valley excitons into quantum open-system frameworks, continue to be warmly remembered.  This work was supported by the National Natural Science Foundation of China (Grants No.~12674075, No.~12274256, No.~11874236, No.~11974047, No.~12022413, No. 92265104, No.~11674331, and No.~12504215), the Major Basic Program of Natural Science Foundation of Shandong Province (Grant No.~ZR2021ZD01), the Natural Science Foundation of Shandong Province (No. ZR2025QC1495), the National Key R\&D Program of China (Grants No. 2022YFA1403200), and the ``Strategic Priority Research Program (B)'' of the Chinese Academy of Sciences (Grant No.~XDB33030100).

%

\end{document}